\documentclass{aa}  

\usepackage{graphicx}
\usepackage{txfonts}
\usepackage{natbib}
\usepackage{lipsum}
\usepackage[bbgreekl]{mathbbol}

\bibpunct{(}{)}{;}{a}{}{,} 

\usepackage{color, listings, setspace, courier, relsize}

\definecolor{Code}{rgb}{0,0,0}
\definecolor{Decorators}{rgb}{0.5,0.5,0.5}
\definecolor{Numbers}{rgb}{0.5,0,0}
\definecolor{MatchingBrackets}{rgb}{0.25,0.5,0.5}
\definecolor{Keywords}{rgb}{0,0,1}
\definecolor{self}{rgb}{0,0,0}
\definecolor{Strings}{rgb}{0,0.63,0}
\definecolor{Comments}{rgb}{0,0.63,1}
\definecolor{Backquotes}{rgb}{0,0,0}
\definecolor{Classname}{rgb}{0,0,0}
\definecolor{FunctionName}{rgb}{0,0,0}
\definecolor{Operators}{rgb}{0,0,0}
\definecolor{Background}{rgb}{0.98,0.98,0.98}
\definecolor{Booleans}{rgb}{0.572,0,0.572}
\definecolor{BuiltinFunction}{rgb}{0.572,0,0.572}
\definecolor{BuiltinConstant}{rgb}{0.572,0,0.572}
\definecolor{Asterisk}{rgb}{0.670,0,1}

\lstdefinelanguage{Python}{
    	numbers=left,
    	numberstyle=\footnotesize,
    	numbersep=7pt,
    	xleftmargin=1.26em,
    	framextopmargin=2em,
    	framexbottommargin=2em,
    	showspaces=false,
    	showtabs=false,
    	showstringspaces=false,
    	frame=l,
    	tabsize=4,
    	stepnumber=1,
	basicstyle=\ttfamily\small,
    	backgroundcolor=\color{Background},
    	breaklines=True,
    	postbreak=\mbox{\textcolor{red}{$\hookrightarrow$}\space},
	commentstyle=\color{green}\ttfamily,
    	stringstyle=\ttfamily\color{Strings},
    	morecomment=[s][\color{Strings}]{'}{'}, 
    	stringstyle=\ttfamily\color{Comments},
    	morecomment=[s][\color{Comments}]{\#}{\#}, 			
	stringstyle=\ttfamily\color{Strings},
	morekeywords={import,from,class,def,for,while,if,is,in,elif,else,not,and,or,print,break,continue,return,access,as,except,exec,finally,global,import,lambda,pass,print,raise,try,assert},
    	keywordstyle={\color{Keywords}\bfseries}, 
    	morekeywords={[2]True,False,None},
    	keywordstyle={[2]\color{BuiltinConstant}\slshape},
	otherkeywords={[2]*},
	keywordstyle={[2]\color{Asterisk}},
	emph={self},
	emphstyle={\color{self}\slshape}	
}
   
\lstdefinelanguage{bash}{
    numbers=none,
    numberstyle=\footnotesize,
    numbersep=7pt,
    xleftmargin=1.26em,
    framextopmargin=2em,
    framexbottommargin=2em,
    showspaces=false,
    showtabs=false,
    showstringspaces=false,
    frame=none,
    tabsize=4,
    stepnumber=1,
    basicstyle=\ttfamily\small\setstretch{1},
    backgroundcolor=\color{Background},
    breaklines=True,
    postbreak=\mbox{\textcolor{red}{$\hookrightarrow$}\space},
}    

\newsavebox{\LstBox}

\newcommand{\bt}{\boldsymbol{\theta}}
\newcommand{\bvt}{\boldsymbol{\vartheta}}
\newcommand{\btt}{\tilde{\boldsymbol{\theta}}}

\newcommand{\tE}{\theta_{\text{\scalebox{.8}{E}}}}

\newcommand{\ba}{\boldsymbol{\alpha}}
\newcommand{\bha}{\hat{\ba}}
\newcommand{\bta}{\tilde{\ba}}
\newcommand{\ha}{\hat{\alpha}}

\newcommand{\kp}{\kappa}
\newcommand{\hkp}{\hat{\kp}}
\newcommand{\tkp}{\tilde{\kp}}

\newcommand{\bb}{\boldsymbol{\beta}}
\newcommand{\bhb}{\hat{\bb}}
\newcommand{\hb}{\hat{\beta}}

\newcommand{\hpsi}{\hat{\psi}}
\newcommand{\tpsi}{\tilde{\psi}}

\newcommand{\bn}{\boldsymbol{\nabla}}
\newcommand{\sub}[2]{#1_{\text{\scalebox{.9}{#2}}}}

\newcommand{\hA}{\hat{\mathcal{A}}}

\newcommand{\pythonpackage}{\texttt{pySPT}}
\newcommand{\pythonpackageLink}{\texttt{\small https://github.com/owertz/pySPT}}
\newcommand{\tutorialLink}{\texttt{\small https://github.com/owertz/pySPT\_tutorials}}

\newcommand{\Python}{\texttt{python}}
\newcommand{\pythonPSL}{\texttt{https://docs.python.org/2/library/index.html}}
\newcommand{\PEPeight}{\texttt{\small https://www.python.org/dev/peps/pep-0008/}}

\newcommand{\spt}{\texttt{spt}}

\newcommand{\sm}{\texttt{sourcemapping}}
\newcommand{\SM}{\texttt{SourceMapping}}
\newcommand{\lensing}{\texttt{lensing}}
\newcommand{\Model}{\texttt{Model}}

\newcommand{\sigmaP}{\sigma^{\text{\scalebox{.9}{P}}}}

\begin{document} 

\title{\pythonpackage: a package dedicated to the source position transformation}

\titlerunning{\pythonpackage}

\author{Olivier Wertz \and Bastian Orthen}       

\institute{Argelander-Institut f\"ur Astronomie, Universit\"at Bonn, Auf dem H\"ugel 71, 53121 Bonn, Germany}

\date{Received November 6th, 2017; accepted ??? ?, 2017}

\abstract
{The modern time-delay cosmography aims to infer the cosmological parameters with a competitive precision from observing a multiply imaged quasar.
The success of this technique relies upon a robust modeling of the lens mass distribution. Unfortunately strong degeneracies between density profiles that 
lead to almost the same lensing observables may bias precise estimate for the Hubble constant. The source position transformation (SPT), which covers 
the well-known mass sheet transformation (MST) as a special case, defines a new framework to investigate these degeneracies. 
In this paper, we present \pythonpackage, a \Python\ package dedicated to the SPT. 
We describe how it can be used to evaluate the impact of the SPT on lensing observables. 
We review most of its capabilities and elaborate on key features that we used in a companion paper regarding SPT and time delays.
\pythonpackage\ also comes with a sub-package dedicated to simple lens modeling. It can be used to generate lensing related quantities for a wide variety of lens models,
independently from any SPT analysis.
As a first practical application, we present a correction to the first estimate of the impact on time delays of the SPT, which has been experimentally found in \citet{SPT_SS13} between a softened power-law 
and a composite (baryons + dark matter) lenses. 
We find that the large deviations predicted in \citet{SPT_SS14} have 
been overestimated due to a minor bug (now fixed) in the public lens modeling code \texttt{lensmodel} (v1.99). 
We conclude that the predictions for the Hubble constant deviate by $\sim 7$\%, first and foremost caused by an MST. 
The latest version of \pythonpackage\ is available at \pythonpackageLink. We also provide tutorials to describe in detail how making best use of \pythonpackage\ at \tutorialLink. 
%
%
}




\maketitle
\titlerunning
\authorrunning

\section{Introduction}
\label{section:introduction}
For about a decade, the modern time-delay cosmography, namely the cosmological parameter inferences from time delay measurements in strong gravitational lensing, have been achieved with 
an increasingly competitive precision \citep[for a recent review, see][ and references therein]{Treu_Marshall_TD_Cosmography_review_2016}. 
A crucial step of this technique relies upon a robust characterization of the gravitational potential which produces the multiply imaged configuration of a background 
bright quasar \citep[see, e.g.,][]{Keeton_substructures_2003, Fassnacht_B1608p656_2006, Suyu_2010_B1608+656_H0, Wong_2011, Suyu_timeDelayDistance_2013, Wong_HOLiCOW_2017}. 
This gravitational potential is essentially produced by a main deflector but also by any mass distributions lying along the line of sight to the source \citep[][]{Seljak_1994, BarKana_1996}. 
Unfortunately, modeling the main lens mass distribution faces a major hurdle, namely the existence of degeneracies between plausible lens density profiles. In fact, a significant freedom exists in choosing lens models
that produce the same image configurations but predict different products of time delays and Hubble constant, $H_0\,\Delta t$ \citep[][]{SPT_SS13}. Thus, these degeneracies translate into systematic errors that 
propagate to $H_0$.

Now well-known to the lensing community, the first lensing invariance to have been pointed out is the mass-sheet degeneracy \citep[MSD,][]{Falco_MST_1985}. 
A dimensionless surface mass density $\kappa(\bt)$ and all the modified $\kappa_{\lambda}(\bt)$ under the mass-sheet transformation (MST) defined as
 \begin{equation}
	\kappa_{\lambda}(\bt) = \lambda\,\kappa(\bt) + (1 - \lambda) \ ,
	\label{MST}
\end{equation}
along with the corresponding unobservable source rescaling $\bb \rightarrow \lambda\,\bb$, lead to identical lensing observables, except for the time delays which transform like 
$\Delta t \rightarrow \lambda\,\Delta t $. This pure mathematical degeneracy has nothing to do with the gravitational perturbations caused by external masses 
along the line of sight. 
Whereas different solutions have been already proposed to reduce its impact on time-delay cosmography \citep[see, e.g., \S 3 in][and references therein]{Treu_Marshall_TD_Cosmography_review_2016}, 
none of them succeed in unambiguously breaking the MSD. The source position transformation (SPT), a more general class of degeneracies which includes 
the MST as a special case, has been introduced in \cite{SPT_SS14} and carried forward in \citet[][]{SPT_USS17}. The SPT defines a new mathematical framework that includes degeneracies that 
have been neither described nor considered in time-delay cosmography before. The first estimation of its impact on time delays is given in \cite{SPT_SS14} where the authors show experimentally that 
predictions for $H_0$ can deviate by $\sim 20\%$.

Recently, a detailed analysis of how the SPT may affect the time-delay cosmography has been presented in \citet[][]{SPT_WOS17}. 
To address this question, we started by developing a flexible numerical framework that encompasses well-tested and efficient implementations of most of the analytical results published in \cite{SPT_SS14} 
and \citet[][]{SPT_USS17}. Numerous additional features were then added, giving rise to \pythonpackage, an easy-to-use and well-documented \Python\ package dedicated to the SPT. 
We used \pythonpackage\ to draw the conclusions presented in \citet[][]{SPT_WOS17}. 
Thus, our package is released as open-source, making our results easy to reproduce. 
Beyond that, we also hope that it will be useful to the time-delay cosmography community to quantify the systematic errors that are introduced by the SPT when inferring $H_0$.

This paper is organized as follows. For readers who are not familiar with the SPT, we outline its basic principles in Sect.\,\ref{section:sptPrinciple}. 
Section\,\ref{section:overview} gives an overall description of the package design and features, whereas we dive into the details in Sect.\,\ref{section:subpackages}. 
In Sect.\,\ref{section:TD}, we present the corrected version of some results presented in \citet{SPT_SS14}. We summarize our
findings in Sect.\,\ref{section:conclusions}

\section{The principle of the source position transformation}
\label{section:sptPrinciple}


This section focuses on the basic idea underlying the SPT and its mathematical framework. For a detailed discussion the reader is referred to \cite{SPT_SS14} and \citet[][]{SPT_USS17}. 
All over this paper, we adopt the standard gravitational lensing notation \citep[see ][]{Schneider_Saas-Fee}.

The relative lensed image positions $\bt_i(\bt_1)$ of a background point-like source located at the unobservable position $\bb$ constitute the lensing observables that we measure with the highest accuracy and precision. 
As a typical example, just a few mas can be achieved with deep HST observations. When $n$ images are observed, the mapping $\bt_i(\bt_1)$ only provides the constraints 
\begin{equation}
	\bt_i - \ba(\bt_i) = \bt_j - \ba(\bt_j) \ , \qquad \qquad \forall\, 1 \leq i < j \leq n \ ,
	\label{spt_constraints}
\end{equation}
where $\ba(\bt)$ corresponds to the deflection law caused by a foreground surface mass density $\kappa(\bt)$, the so-called lens. 
The SPT addresses the following question: can we define an alternative deflection law, denoted as $\bha(\bt)$, that preserves the mapping $\bt_{i}(\bt_{1})$ for a unique source?  
If such a deflection law exists, the alternative source position $\bhb$ differs in general from $\bb$. Furthermore, it defines the new lens mapping $\bhb = \bt - \bha(\bt)$, which leads to
\begin{equation}
	\bt = \bb + \ba(\bt) = \bhb + \bha(\bt) \ .
	\label{spt_implicit}
\end{equation}
An SPT consists in a global transformation of the source plane formally defined by a
one-to-one mapping $\bhb(\bb)$, unrelated to any physical contribution such as the external convergence \citep{SPT_SS13}.
To preserve the mapping $\bt_{i}(\bt_{1})$, the alternative deflection law thus reads 
\begin{equation}
	\bha(\bt) = \ba(\bt) + \bb - \bhb(\bb) = \ba(\bt) + \bb - \bhb(\bt - \ba(\bt)) \ , 
	\label{hat_alpha_definition}
\end{equation}
where in the first step we used Eq.\,\eqref{spt_implicit} and in the last step we inserted the original lens equation. 
As defined, the deflection laws $\ba(\bt)$ and $\bha(\bt)$ yield exactly the same image positions of the source $\bb$ and $\bhb$, respectively. 

Because $\bha$ is in general not a curl-free field, it cannot be expressed as the gradient of a deflection potential caused by a mass distribution $\hkp$.
Provided its curl component is sufficiently small,  \citet[][]{SPT_USS17} have established that one can find a curl-free deflection law $\bta = \bn \tpsi$ that 
is similar to $\bha$ in a finite region. In other words, $\bta$ yields the same sets of multiple images up to the astrometric accuracy $\sub{\varepsilon}{acc}$ of current observations. 
Two image configurations are considered indistinguishable when they satisfy  
\begin{equation}
	|\Delta \bt| \coloneqq \left|\btt - \bt\right| < \sub{\varepsilon}{acc} \ ,
	\label{astrometricCriterion}
\end{equation}
for all images $\bt$ of the source $\bb$, and corresponding images $\btt$ of the source $\bhb$ with $\btt = \bhb + \bta(\btt)$.
In \citet[][]{SPT_USS17}, the adopted similarity criterion reads
\begin{equation}
	 |\Delta \ba(\bt)| \coloneqq |\bta(\bt) - \bha(\bt)| < \sub{\varepsilon}{acc} \ ,
	\label{criterion}
\end{equation}
in a finite region of the lens plane denoted as $\mathcal{U}$ where multiple images occur. 
Even though this criterion cannot actually guarantee $|\Delta \bt| < \sub{\varepsilon}{acc}$ 
over $\mathcal{U}$, there exists in general a subregion $\mathcal{U}' \subset \mathcal{U}$ that includes image configurations for which Eq.\,\eqref{astrometricCriterion} is 
satisfied \citep[see \S 4.1 in][]{SPT_WOS17}. Thus, an SPT $\bhb(\bb)$ leads to an alternative lens profile $\hkp$ which gives rise to the SPT-transformed deflection 
law $\bha$, then its curl-free counterpart $\bta$ is defined based upon the criterion \eqref{criterion}, and may lead to indistinguishable image configurations produced by $\ba$. 
To conclude this section, we note that the deflection law $\bta$ is produced by a surface mass density $\tkp \coloneqq \bn \cdot \bta / 2$ that equals $\hkp$ by construction. 






\section{Package overview}
\label{section:overview}

\begin{lrbox}{\LstBox}
\begin{lstlisting}[language=bash]
git clone --recursive https://github.com/owertz/pySPT
\end{lstlisting}
\end{lrbox}

\pythonpackage\ is being developed in \Python\ \citep{python} and only relies on packages included in the \Python\ standard library\footnote{\pythonPSL} and the proven open-source 
libraries \texttt{numpy} \citep{numpy}, \texttt{scipy} \citep{scipy}, and \texttt{matplotlib}  \citep{matplotlib}.
The code design and development follow effective practices for scientific computing such as using test cases and a profiler to identify bugs and bottlenecks, keeping an effective collaboration 
thanks to a version control system (\texttt{git}), and providing an extensive documentation \citep[][]{Wilson_2014}.
This open source code is available on \texttt{GitHub}\footnote{\pythonpackageLink. For \texttt{git} users, \pythonpackage\ can be cloned directly from the source code repository by using the following bash command \\ 
\usebox{\LstBox}
}
and comes along with a clear description on how to install it. To make the best use out of \pythonpackage, a quick start guide and several tutorials are also provided 
in the form of \texttt{Jupyter} notebooks.
Benefiting from the \Python\ object-oriented paradigm, the structure of \pythonpackage\ is highly modular, which avoids `code clones' and makes it less sensitive to bug propagation. 

The code is composed of several modules build from various classes and is organized in a dozen of sub-packages. Its core is separated into three main sub-packages, 
referred to as \lensing, \sm, and \spt. The reason for that is simple.
The mapping $\bhb(\bb)$ presented in Sec.\,\ref{section:sptPrinciple} defines a one-to-one connection between the source plane and its SPT-transformed counterpart. 
Through the lens equation $\bb = \bt - \ba(\bt)$, the deflection law (arising from a lens model) characterizes the link between the source plane and the image plane. 
Thus, it is only together with $\ba(\bt)$ that $\bhb(\bb)$ gives rise to the alternative deflection law $\bha(\bt)$ defined in Eq.\,\ref{hat_alpha_definition}. 
As a result, the most significant \pythonpackage\ sub-packages are \lensing\  to basically deal with $\ba(\bt)$, \sm\  to define $\bhb(\bb)$, 
and \spt\ to describe $\bha(\bt)$ (and all the SPT-transformed lensing quantities).
In the remainder of this section, we describe briefly their main functionalities.\footnote{We adopt the naming conventions 
advocated in the Python Enhancement Proposals 8 (PEP-8). In particular, sub-packages have all-lowercase names while class names use the so-called CapWords
convention. The PEP-8 is accessible at \PEPeight}

The sub-package \lensing\ shares a lot of functionalities with other lensing-dedicated softwares such as \texttt{gravlens} \citep{Keeton_computationalMethods_2001}.
From its class \Model, we generate a lens model object that allows us to perform a wide range of basic lensing calculations. 
Strictly speaking, this part of the code is not related to the SPT and it can be used independently of any SPT analysis. 
Nevertheless, we have decided to implement this part of the code for a simple reason: as a built-in feature of \pythonpackage, 
the class \Model\  provides to its instances the adequate structure that is required to match the \spt\ requirements. This 
makes \lensing\  
more convenient to use 
than a third party code. 
The sub-package \sm\ is used to define an SPT $\bhb(\bb)$. Several forms of $\bhb(\bb)$ are implemented, such as particular radial stretchings presented in \citet{SPT_SS14}. 
Moreover, the code is designed to accept any user-defined SPT. 
With \sm\  comes also functionalities 
to characterize the mapping $\bhb(\bb)$, such as testing whether it is one-to-one over a given region. More details are given in Sect.\,\ref{subsection:sm}.
The sub-package \spt\ is the heart of \pythonpackage. It is designed to work alongside with \sm\  and \lensing\  in order to provide all the 
basic SPT-transformed quantities, such as $\hkp$, $\bha$, $\hpsi$, $\hat{\mathcal{A}}$, $\btt$, $\bta$, $\tpsi$. One can also derive the SPT-transformed time delays $\Delta \hat{t}$ and $\Delta \tilde{t}$, and most of the quantities presented in \citet[][]{SPT_SS14}, \citet[][]{SPT_USS17}, and \citet[][]{SPT_WOS17}. 
This makes the results presented in these papers straightforward to reproduce. A detailed description of the tools provided by \spt\ is given in Sect.\,\ref{subsection:spt}. 

\pythonpackage\ also includes several sub-packages dedicated to specific tasks. Based on the package \texttt{multiprocessing} of the standard library, \texttt{multiproc} provides an efficient tool to 
parallelize functions and methods in a straightforward way. As such, most of the \pythonpackage\ features support parallel computing to fully leverage multiple processors. 
\texttt{grid} helps us to create different types of mesh grids. These are of practical interest for generating maps of lensing quantities in a particular region. 
To calculate $\bta$ efficiently, we follow \citet[][]{SPT_USS17} that suggests using a Riemann mapping to handle a pole numerically. Thus, \pythonpackage\ contains the 
sub-package \texttt{integrals} that includes functionalities to deal with and to illustrate conformal mappings.


\section{Analyzing the impact of the SPT with \pythonpackage}
\label{section:subpackages}


\subsection{The sub-package \lensing\  to deal with lens models}
\label{subsection:lensing}

\begin{figure*}
	\centering
   	\includegraphics[width=\hsize]{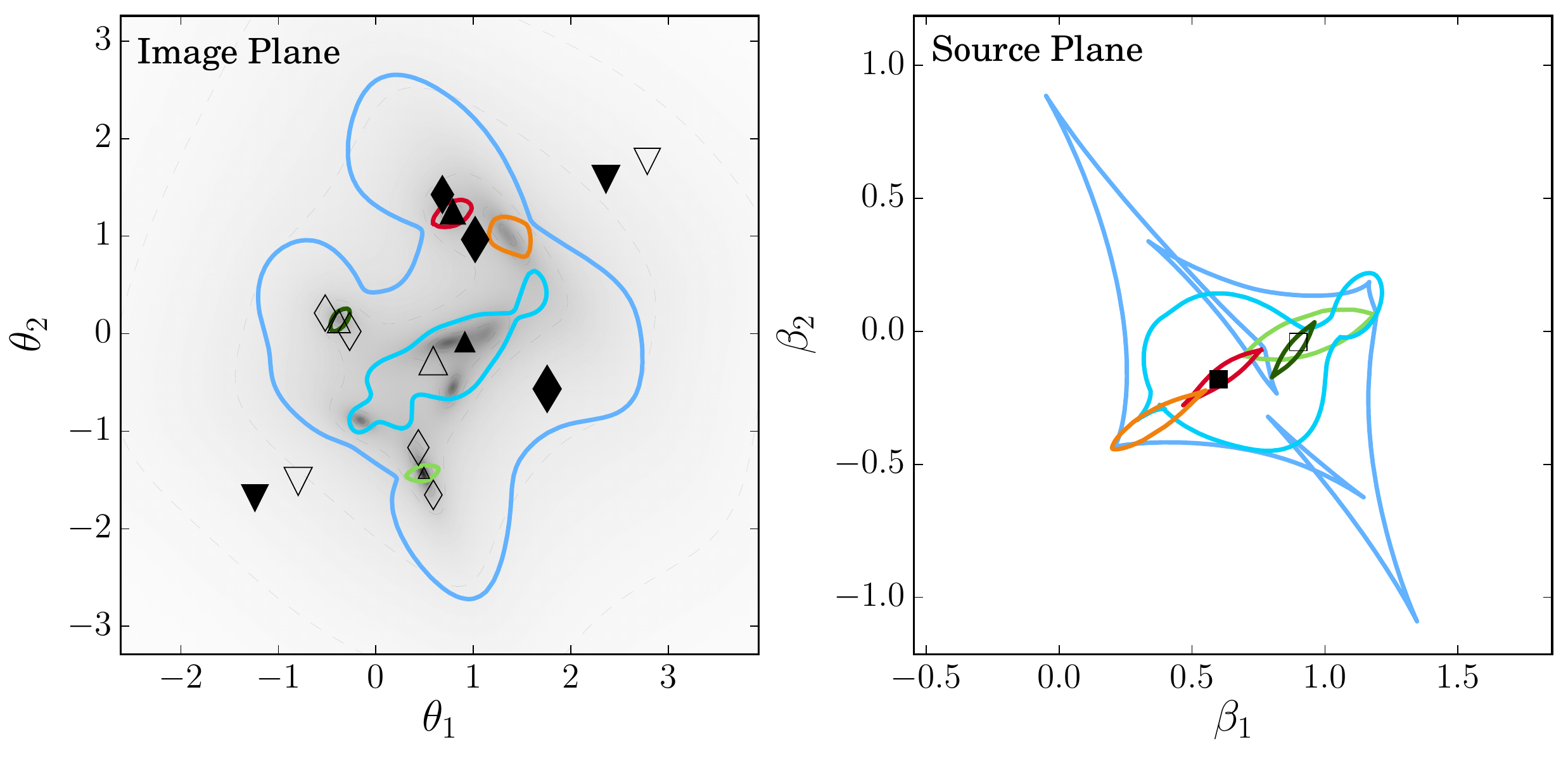}
     	\caption{Example illustrating some capabilities of the sub-package \lensing. 
	\emph{Left}: eight lens profiles with different parameters are combined to generate a complex mass distribution. The total
	surface mass density is shown in tones of grey and the dashed curves highlight few particular iso-density contours. The colored thick curves show the critical curves and the filled and empty markers locate
	the lensed image positions of two different sources shown in the right panel. The inverted triangles correspond to images of type I (maximum of $\tau$), the diamonds to type II 
	(saddle point of $\tau$), and triangles to type III (maximum of $\tau$). The size of the markers is log-proportional to the magnification of the images. 
	\emph{Right}: the colored lines show the caustics and the two square locate the sources. The filled (resp. empty) square has seven (resp. nine) images, all shown in the left panel. The axis scale is arcseconds
	but unit of $\theta_{\text{E}}$ can also be used.}
       	\label{figure:MODEL_example}       
\end{figure*}   
 
   


To analyze how the SPT may alter lensing observables in a quantitative way, we need to select a model that characterizes a mass distribution.
A range of lensing observables are then generated and compared against those produced by an SPT-transformed version of the original model.
The sub-package \lensing\  relies on the class \Model\  whose one of its instance defines a lens model and provides efficient tools to compute a wide range of lensing quantities. 
Most of the mass profiles described in \citet{Keeton_MassCatalog_2001} are available in the sub-package \texttt{catalog}, together with a complete documentation.
A brief aside here to note that, to our knowledge, no analytical expression of the deflection potential for the generalized pseudo-NFW can be found in the literature. Hence, we have derived
this expression and provide the result in Appendix\,\ref{appendix:massmodels}.
The class \Model\  is also designed to accept a user-defined lens model, as long as some conventions are respected. 
At least, the deflection potential $\psi(\bt)$ must be provided as either a \Python\ function or a \texttt{C} shared library. 
Those that are not defined among $\ba$, $\kappa$, and $\partial \ba/\partial \bt$, are then computed from numerical approximation at the expense of a more intensive usage of computer resources.
Thanks to operator overloading\footnote{Operator overloading is a special case of polymorphism that is well defined in the object-oriented programming (OOP) paradigm. 
Thus, this feature is not only a \Python\ syntactic sugar but exists in any other OOP languages.}, 
arbitrarily complicated composite models can be obtained by simply adding several \Model\  instances. 
For example, this feature is used to generate quadrupole models, namely the combination of an axisymmetric matter distribution plus some external shear. 
To go a step further, the axisymmetric part may itself be composed of different components, whereas additional contributions can be included at arbitrary positions. 


The computational methods implemented with the class \Model\  follow the prescription of \citet{Keeton_computationalMethods_2001}. In particular, the so-called `tiling' algorithm 
(with adaptive grid) is used to solve the lens equation 
and locate the critical curves.
Other fundamental lensing quantities are available, such as the caustics, the basic image properties (image type, magnification factor, parity, odd-number and magnification theorem checks, ...), 
the Fermat potential $\tau(\bt)$, and time delays between image pairs $\Delta t(\bt_i, \bt_j) \equiv \Delta t_{ij}$. 
In Fig.\,\ref{figure:MODEL_example}, we illustrate the capabilities of the sub-package \lensing\  by showing several lensing quantities produced by a complex mass distribution. 
The workflow for generating data used in this figure is as follows\footnote{A \texttt{Jupyter} notebook dedicated to this figure is available in the \pythonpackage\ repository on \texttt{GitHub}. 
Besides details about the workflow shown above, it includes the code we used to plot Fig.\,\ref{figure:MODEL_example}.}:

The sub-package \lensing\  also includes \texttt{C} shared libraries that implement $\psi$, $\ba$, $\kappa$, $\partial \ba / \partial \bt$ for all the lens models. 
As shown in Sect.\,\ref{subsection:spt}, the curl-free deflection angle $\bta$ is obtained by means of line and surface integrals of functions 
that involve $\bha$ (see Eq.\,\ref{hat_alpha_definition}) and $\hkp = \bn \cdot \bha / 2$ \citep{SPT_USS17}. Thus, the original deflection angle $\ba$ is evaluated as many 
times as the number of iterations required by the solver to converge. Even though this procedure is efficient when only one $\bta$ is evaluated, high performance
becomes critical when $\bta$ needs to be evaluated on a dense grid. Thanks to the foreign function interface module \texttt{ctypes}\footnote{\texttt{ctypes} is included 
in the \Python\ standard library: \texttt{\small https://docs.python.org/2.7/library/ctypes.html}}, the use of the \texttt{C} shared libraries speeds up significantly the 
execution of pure \Python\ code. \pythonpackage\ comes along with a tutorial in the form of a \texttt{Jupyter} notebook that describes in details how to deal with \texttt{C} shared libraries.


\subsection{Defining $\bhb(\bb)$ with the sub-package \sm\ }
\label{subsection:sm}


Now that we have defined a lens model, we choose an SPT by defining a source mapping $\bhb \equiv \bhb(\bb)$. 
To each position $\bb$ of the source plane, this mapping associates a new and unique position $\bhb$ in the source plane.
The so-called radial stretching is simply defined as
\begin{equation}
	\bhb(\bb) = \left[1 + f(|\bb|) \right] \bb \ ,
	\label{radial_stretching}
\end{equation}
where $f$ is called the deformation function. With $\bhb$ defined this way, $\bhb(\bb_j)$ always lies on the line 
passing through $\boldsymbol{0}$ and $\bb_j$. The most simple case of radial stretching corresponds to a constant deformation function, 
$f(|\bb|) = \lambda - 1$ with $\lambda \in \mathbb{R}$, which leads to the well-known MST, $\bhb = \lambda\,\bb$.
In \citet{SPT_WOS17}, we focus most of the work on the radial stretching \eqref{radial_stretching} where the deformation function 
$f(|\bb|)$ is defined as the lowest-order expansion of more general functions
\begin{equation}
	f(|\bb|) = f_0 + \frac{f_2}{2 \tE^2} |\bb|^2 \ ,
	\label{deformation_function}
\end{equation}
where $f_0 \coloneqq f(0)$, $f_2 \coloneqq \tE^2\,f''(0)$, $\tE$ is the Einstein angular radius, and $f$ is an even function of $|\bb|$ to preserve 
the symmetry \citep{SPT_SS14}. When $f_2 = 0$, this case simplifies to a pure MST with $\lambda = 1 + f_0$. 
In Table\,\ref{table:lensmapping}, we provide a list of the radial stretchings implemented in \sm. 
The rationale behind the choice of these particular deformation functions are motivated in \citet{SPT_SS14}. 
The first column refers to the \texttt{id} used to identify which source mapping one wants to select. A specific example is given below.
Besides defining an SPT, \sm\  also includes a few functionalities for characterizing the mapping $\bhb(\bb)$. 
In particular, one can test whether the source mapping is one-to-one over a given region, compute the Jacobi matrix 
$\mathcal{B}(\bb) = \partial \bhb(\bb) / \partial \bb$, and provide its decomposition $\mathcal{B}(\bb) = B_1 \mathcal{I} + B_2 \Gamma(\eta)$ that is useful to evaluate the amplitude
of the curl component of $\bha$ \citep[see \S 4.1 in][]{SPT_SS14}.
As an example, the code below illustrates how to define and characterize an MST with $\lambda = 1 (\equiv 1 + f_0)$: 
\begin{lstlisting}[language=Python]
from pySPT.sourcemapping import SourceMapping
from numpy.random import uniform

# Define an MST with lambda=1. : #
sm = SourceMapping(1, f0=0., f2=0.)

# For a random source position beta ...#
beta = uniform(-1,1,2)

# ... we compute some quantities #
hatbeta = sm.modified_source_position(beta)
jm = sm.jacobi_matrix_spt_mapping(beta)
B1,B2 = sm.B(beta)
eta = sm.phase_shear(beta)
\end{lstlisting} 
Similarly, we can first define the deformation function $f$ that characterizes an MST with $\lambda$ as unique argument, and pass it to \SM\  as an argument.
For efficiency, the first derivative $f'$ of the deformation function with respect to $|\bb|$ can also be passed\footnote{For axisymmetric profile $\kappa(\theta) = \kappa(|\bt|)$, the analytical 
expression for $\hkp$ involves $f'$ \citep[see equation 16 in][]{SPT_SS14}.}. This process illustrates how we can work with a user-defined SPT:
\begin{lstlisting}[language=Python]
from pySPT.sourcemapping import SourceMapping

# Define the deformation function ... # 
def f(b, f_args):
    return f_args[0] - 1

# ... and its first derivative #
def df(b, f_args):
    return 0.0

# Define the MST with lambda=1 #
f_args = [1.0]
sm = SourceMapping(0, f, f_args, df)
\end{lstlisting}

\begin{table}
	\caption[]{List of radial stretchings implemented in \texttt{lensmapping}}
         \label{table:lensmapping}
      	$$ 
         \begin{array}{p{0.1 \linewidth}p{0.5 \linewidth}l}
         	\hline
            	\noalign{\smallskip}
            	\texttt{id}	&	$f(|\bb|)$ & $\text{Arguments}$ \\
            	\noalign{\smallskip}
            	\hline
            	\noalign{\smallskip}
             	1	&	$f_0 + f_2 |\bb|^2 / 2$  									&  f_0, f_2 \\
		    	2	&	$f_0 + \beta_0^2\,f_2\,|\bb|^2 \ \left[2\,\left(\beta_0^2 + |\bb|^2\right) \right]$ 		& f_0, f_2, \beta_0		\\
	   		3	&	$2 f_0 / \cosh{\left(|\bb| / \beta_0\right)} - f_0 \ , \text{with}$					& f_0, \tE		\\
				&	$\beta_0 = \tE \sqrt{3 (1-f_0)/(1+f_0)}$										&		\\
			4	&	$a\left[1 - \cos{(c\,|\bb|)} \right]$											& a, c		\\
            	\noalign{\smallskip}
            	\hline
         \end{array}
         $$
     	 
	\tablefoot{The first column \texttt{id} is used to select the source mapping when we instantiate the class \SM\  (see in text for more details).}     
\end{table}




\subsection{Deriving SPT-transformed quantities with \spt }
\label{subsection:spt}

\begin{figure*}
	\centering
   	\includegraphics[width=0.5 \linewidth]{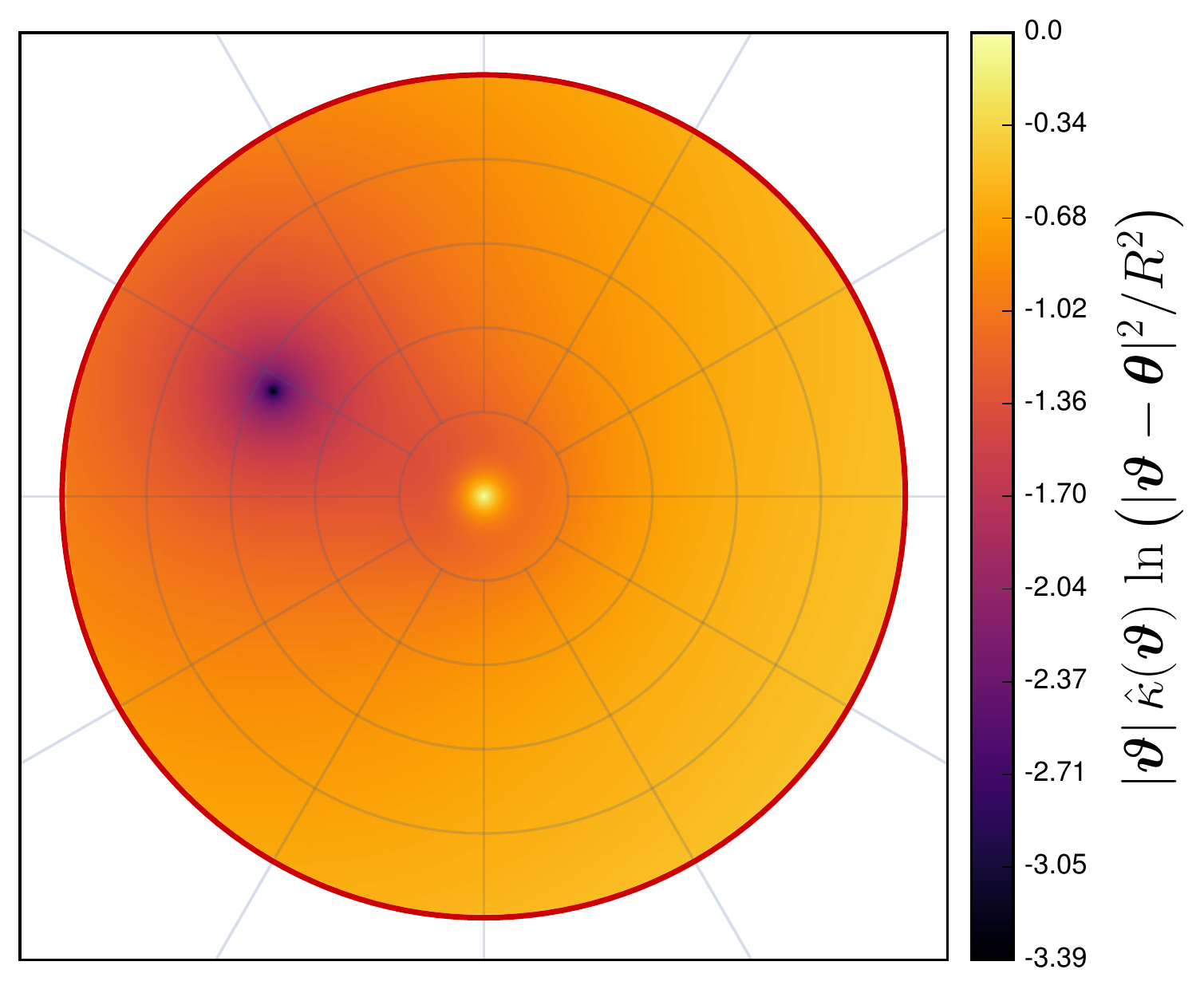}
   	\includegraphics[width=0.408 \linewidth]{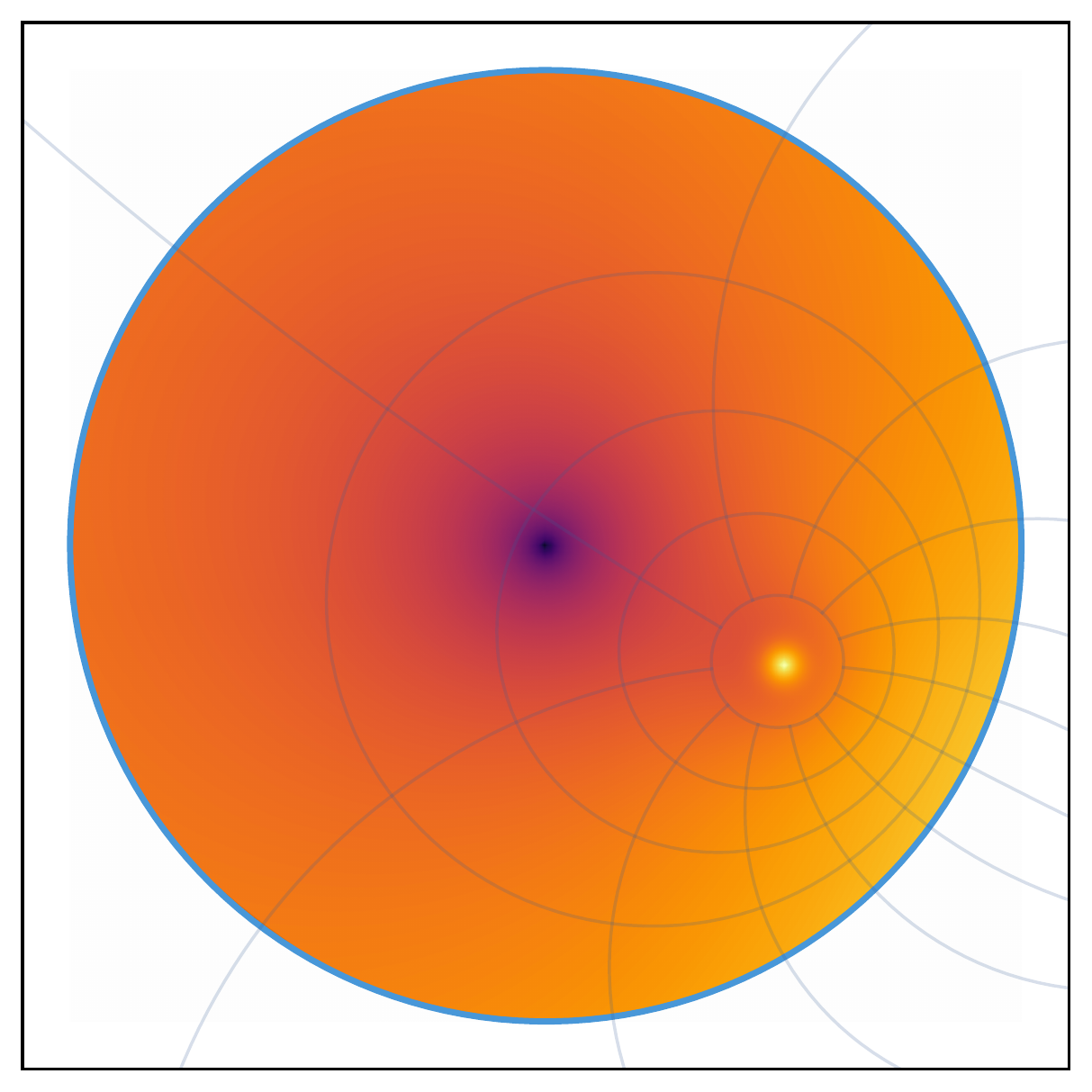}	
     	\caption{Representative example that illustrates how the integration domain $\mathcal{U}$ is mapped onto the unit disk in the complex plane under a Riemann mapping.
		\emph{Left}:	the color-coding depicts the integrand $|\bvt|\,\hkp(\bvt)\,\ln(|\bvt - \bt|^2 / R^2)$ for all $\bvt \in \mathcal{U}$. It shows a pole at $\bvt = \bt$ with $\bt = (-0.5,0.25)\,\theta_{\text{E}}$ and a secondary peak at the origin caused
					by $\kappa(|\bvt| = 0)$. The lens model is an NIS $(\theta_{\text{c}} = 0.1\theta_{\text{E}})$ plus external shear $(\gamma_{\text{p}} = 0.1)$ transformed by the radial
					stretching $\bhb(\bb) = f_0 + f_2 |\bb|^2 / (2 \theta_{\text{E}}^2)$ with $(f_0,f_2) = (0,0.55)$. The red circle delimits $\mathcal{U}$ with radius $R$.
		\emph{Right}: 	integrand after applying the Riemann mapping described in Appendix A in \citet{SPT_USS17}. The pole $\bvt = \bt$ now lies at the origin of the unit (blue) circle
					of the complex plane. The polar grid (gray lines) helps us to visualize how the Riemann mapping acts on $\mathcal{U}$. 
					For obtaining this figure, we used the sub-package \texttt{integrals} that takes care to deal with both the pole and the second peak.
		}
       	\label{figure:cm}       
\end{figure*}   




In the two previous sections, we have shown how to generate lensing observables produced by a given lens model and how to define an SPT. 
We present here the sub-package \spt, which provides the tools required to determine the SPT-transformed quantities 
$\hkp$, $\bha$, $\hpsi$, $\hat{\mathcal{A}}$, $\Delta \hat{t}$, $\bta$, $\tpsi$, $\btt$, and $\Delta \tilde{t}$.



For any SPT and lens model, the alternative deflection law $\bha$ is implemented as defined in Eq.\,\eqref{hat_alpha_definition}, and the Jacobi matrix of the alternative 
lens equation given in Eq.\,\eqref{spt_implicit} as $\hA(\bt) = (\partial \bhb / \partial \bb) \, (\partial \ba / \partial \bt) \equiv \mathcal{B}(\bb(\bt))\,\mathcal{A}(\bt)$. 
The determination of $\hA$ perfectly illustrates how \spt\ works alongside with \lensing\ and \sm: 
$\mathcal{B}(\bb)$ is obtained with \sm, $\bb(\bt)$ and $\mathcal{A}(\bt)$ are obtained with \lensing, and \spt\ combines the different results to derive $\hA(\bt)$.
%
In the axisymmetric case, $\bha$ is a curl-free field and there exists a deflection potential $\hpsi$ such that 
$\text{d}\hpsi(\theta)/\text{d}\theta = \ha(\theta) = \theta - \hb(\theta - \alpha(\theta))$ where $\theta = |\bt|$.
Thus, $\hpsi$ is obtained as follows
\begin{equation}
	\hpsi(\theta) = \int_{0}^{\theta} \left[ \vartheta - \hb(\vartheta - \alpha(\vartheta)) \right]\,\text{d}\vartheta\ ,
	\label{hat_psi}
\end{equation}
up to a constant independent on $\theta$. Because the integrand in Eq.\,\eqref{hat_psi} depends on the SPT and the lens model choices, no general analytical solution can be derived. The integral is therefore computed numerically using the \Python module \texttt{integrate.quad}\footnote{This package provides an interface to QUADPACK \citep{QUADPACK} whose routines use the adaptive quadrature method to approximate integrals.} 
from \texttt{scipy} \citep{scipy}.
Under a radial stretching, the axisymmetric mass profile $\kappa(\theta)$ transforms into 
$\hkp(\theta) = \kappa(\theta) - [1 - \kappa(\theta)]\,f(\beta(\theta)) - \theta \, \text{det}\,\mathcal{A}(\theta) \, f'(\beta(\theta)) / 2$, where $\beta = |\bb|$ \citep[][]{SPT_SS14}. 
Otherwise, the more general form $\hkp(\bt) = 1 - \text{Tr}(\hA) / 2$ is valid regardless of the lens model symmetry.

When the axisymmetry assumption for the original lens model is dropped, \spt\ also provides the physically meaningful $\bta$ and $\tpsi$.
The analytical expressions implemented in \spt\ are slightly simplified versions of the ones firstly presented in \citet{SPT_USS17}.
The deflection potential $\tpsi$ evaluated at the position $\bt$ in the lens plane explicitly reads
\begin{equation}
	\tpsi(\bt) = \left\langle \tpsi \right\rangle + 2 \int_{\mathcal{U}} H_1(\bt;\bvt)\ \hat{\kp}(\bvt)\ \text{d}^2\vartheta - \int_{\partial\mathcal{U}} H_2(\bt;\bvt)\ \bha \cdot \boldsymbol{n}\ \text{d}s \,
	\label{tilde_psi_simplified}
\end{equation}
where the region $\mathcal{U}$ is a disk of radius $R$, $\left\langle \tpsi \right\rangle$ is the average of $\tpsi$ on $\mathcal{U}$, d$s$ the line element of the boundary curve $\partial\mathcal{U}$, 
\begin{equation}
	H_1(\bt;\bvt) = \frac{1}{4 \pi} \left[\ln\left(\frac{\left|\bvt-\bt\right|^2}{R^2}\right) + \ln\left(1 - \frac{2 \bvt \cdot \bt}{R^2} + \frac{|\bvt|^2 |\bt|^2}{R^4}\right) - \frac{|\bvt|^2}{R^2} \right] \ ,
	\label{H1}
\end{equation}
and
\begin{equation}
	H_2(\bt;\bvt) = \frac{1}{4 \pi} \left[2 \ln\left(\frac{\left|\bvt-\bt\right|^2}{R^2}\right) - 1 \right] \ .
	\label{H2}
\end{equation}
In Appendix \ref{appendix:tildepsi}, we show explicitly that both versions are fully equivalent. 
The corresponding simplified version of the deflection angle $\bta$ can be derived by 
obtaining the gradient of $H_1$ and $H_2$ with respect to $\bt$, which reads 
\begin{eqnarray}
	\bta(\bt) &=& \frac{1}{\pi} \int_{\mathcal{U}} \left(\frac{\bt - \bvt}{|\bt - \bvt|^2} + \frac{|\bvt|^2 \bt - R^2 \bvt}{R^4 - 2 R^2 \bvt \cdot \bt + |\bvt|^2 |\bt|^2}\right)\ \hat{\kp}(\bvt)\ \text{d}^2\vartheta \nonumber \\
	             &-& \frac{1}{\pi} \int_{\partial\mathcal{U}}\frac{\bt - \bvt}{|\bt - \bvt|^2} \ \bha \cdot \boldsymbol{n}\ \text{d}s \ .
	\label{tilde_alpha_simplified}
\end{eqnarray}
To deal with the pole $\bvt = \bt$ in the first term of Eqs.\,\eqref{H1} and \eqref{tilde_alpha_simplified}, we use a Riemann mapping as described in Appendix A in \citet{SPT_USS17} 
and implemented in the sub-package \texttt{integrals}.
This mathematical trick makes the previous integrals easier to solve by mapping $\mathcal{U}$ onto the unit disk in the complex plane, such as the pole is moved at the origin\footnote{The quantity $z$ represents here a complex number $z = x + \mathrm{i}\,y$ with $(x,y) \in \mathbb{R}^2$.} $z = 0$. 
Additional care is however needed in the vicinity of $\bt = \boldsymbol{0}$ where the gradient of $\kappa(\bt)$ may vary significantly and produce a second sharp peak of the integrand. 
This peak may even be a new pole when $\kappa$ is singular at the origin. Thus, physically meaningful lens model should always be favored.
In the left panel in Fig.\,\ref{figure:cm}, we illustrate the integrand\footnote{The term $|\bvt|$ corresponds to the Jacobian of polar coordinates.} 
$|\bvt|\,\hkp(\bvt)\,\ln\left(|\bvt - \bt|^2 / R^2\right)$ of the first term of the integral over $\mathcal{U}$ defined in Eq.\,\eqref{tilde_psi_simplified}. 
For this illustrative example, we choose an NIS $(\theta_{\text{c}} = 0.1\theta_{\text{E}})$ plus external shear $(\gamma_{\text{p}} = 0.1)$ transformed by the radial 
stretching $\bhb(\bb) = f_0 + f_2 |\bb|^2 / (2 \tE^2)$ with $(f_0,f_2) = (0,0.55)$ and $\bt = (-0.5,0.25)\,\tE$. The negative peak comes from the pole $\bvt = \bt$
while the central peak is caused by $\kappa(|\bvt| = 0)$. The right panel in Fig.\,\ref{figure:cm} illustrates the integrand after applying the Riemann mapping. 
The peaks have moved and the pole $\bvt = \bt$ now lies at the origin $z = 0$, as expected. To deal with the second peak, the sub-package \texttt{integrals} takes care to split the 
integration domain to place it on a boundary and to ensure the gradient of the integrand to be smooth in the sub-domains. 

\begin{figure}
	\centering
   	\includegraphics[width=\linewidth]{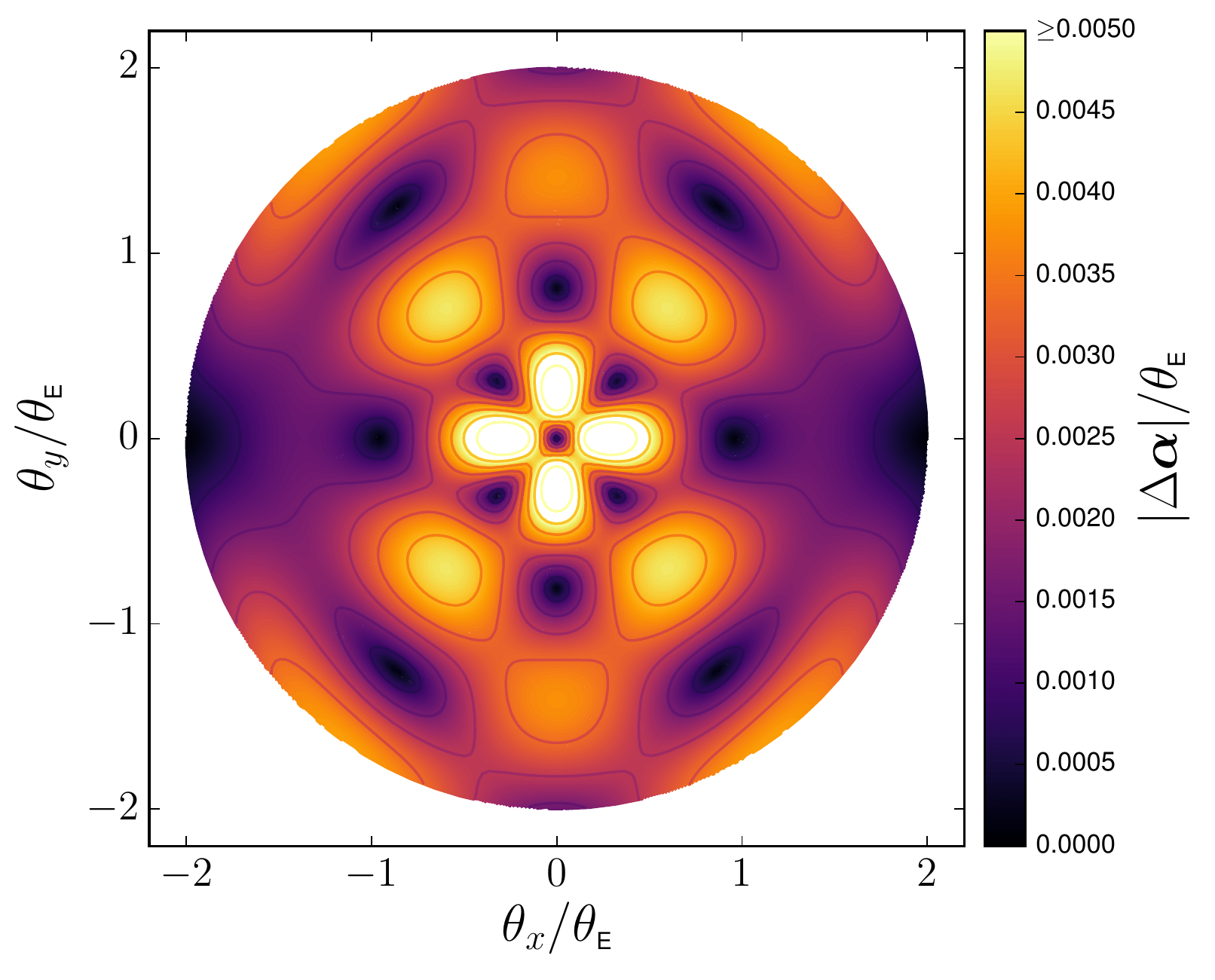}
     	\caption{	Map of $|\Delta \ba(\bt)|$ over a circular grid $|\bt| \leq 2\,\theta_{\text{E}}$ for $f_2 = 0.55$, $\theta_{\text{c}} = 0.1\,\theta_{\text{E}}$ and $\gamma_{\text{p}} = 0.1$. 
			We set the radius $R$ of the circular region $\mathcal{U}$ in such a way that the area of $\mathcal{U}$ is equal to the area of the square grid used in the pure numerical 
			approach, i.e., $R = 4\,\theta_{\text{E}} / \sqrt{\pi} \approx 2.257\,\theta_{\text{E}}$. This figure is similar to the figure $7$ in \cite{SPT_USS17} even though it is based on a different 
			approach (see the text for more details). This figure has been obtained with the sub-package \spt\ in less than five minutes for a grid of about $2 \times 10^4$ positions.
			}
       	\label{figure:Delta_alpha_map_1}       
\end{figure}

With $\bha$ and $\bta$, we can evaluate the SPT validity criterion adopted in \citet{SPT_USS17} and recalled in Eq.\,\eqref{criterion}.
With the same lens model and SPT adopted for Fig.\,\ref{figure:cm}, Fig.\,\ref{figure:Delta_alpha_map_1} shows the map $|\Delta \ba(\bt)|$ over a 
circular grid $|\bt| \leq 2\,\tE$ in the lens plane. It is worth noting that this figure is similar to the map $|\Delta \ba(\bt)|$ illustrated in the figure 7 in \citet{SPT_USS17}
while $\bta$ was obtained from two different and independent approaches.
In \citet{SPT_USS17}, they first calculated $\tpsi$ by solving numerically a Neumann problem thanks to a successive overrelaxation method \citep{Press_1992} on a square 
grid of width $4\,\tE$; then they derived $\bta$ from $\tpsi$ using a second-order accurate finite differencing 
scheme (see their section 3.2 for a detailed overview). This iterative process necessarily requires to systematically calculate $\tpsi$ over the whole square grid. 
Conversely, in Fig.\,\ref{figure:Delta_alpha_map_1}, $\bta$ is obtained directly from the explicit Eq.\,\eqref{tilde_alpha_simplified} 
for each position on a circular sampling grid\footnote{
We may have chosen a square or whatever shape for the sampling grid. The choice of a disk is motivated by the fact that 
the region where multiple images occur is typically $|\bt| \leq 2\,\tE$. 
Furthermore, the radius of the circular sampling grid over which we evaluate $|\Delta \ba|$ and depicted in Fig.\,\ref{figure:Delta_alpha_map_1} is not the radius $R$ of $\mathcal{U}$. 
For each position on the grid, solving the Eq.\,\eqref{tilde_alpha_simplified} requires to define $R$. For consistency, 
we must adopt the same $R$ for each evaluation of $\bta$ which implies that $R$ must be chosen (at least equal or) larger than the radius of the circular sampling grid.
}. 
Thus, although the similarity between the two figures confirms the consistency of the two approaches, the semi-analytical approach implemented in \pythonpackage\ 
yields $\bta$ at a particular position.

In our paper \citet[][]{SPT_WOS17}, we analyzed the impact of the SPT on time delays in details. To achieve this, we compared, for a given lens model, the 
time delays $\Delta t_{ij}$ between image pairs $(\bt_i, \bt_j)$ of a source $\bb$ with the time delays $\Delta \tilde{t}_{ij}$ between the image pairs $(\btt_i, \btt_j)$ of the 
modified source $\bhb$ under an SPT. The images $\btt$ satisfy the lens equation $\bhb = \btt - \bta(\btt) = \btt - \bn \tpsi(\btt)$. By construction of $\bta$, we
expect $\btt_i$ to be close to $\bt$, at least in a subregion $\mathcal{U}' \subset \mathcal{U}$ \citep[see section 4 in][]{SPT_WOS17}. These images $\btt$ 
can be obtained easily thanks to the sub-package \lensing. Because its main class \Model\ is designed to accept a user-defined lens model, we benefit from all the tools provided 
by \lensing\ to characterize the SPT-transformed lens model (see Sect.\,\ref{subsection:lensing}). The code below illustrates how to generate the SPT-transformed lensing quantities
thanks to the use of the sub-packages \lensing, \sm\ and \spt.
\begin{lstlisting}[language=Python, mathescape=true]
from pySPT.spt import SPT
from pySPT.lensing import Model
from pySPT.sourcemapping import SourceMapping

# We start by defining: #
# (1) The lens model #
lens = Model('NIS', p0=(0.1,1.0))            + Model('SHEAR', p0=(0.1,0.0))

# (2) The source mapping #
sm = SourceMapping(1, f0=0.0, f2=0.55)

# (3) The SPT object #
spt = SPT(sm, lens)
spt.load_C_libraries(model_ID='NISG',     sm_ID='IS1')

# (4) The radius of the region $\textcolor{Comments}{\mathcal{U}}$ #
R = 2.257

# (5) The SPT-transformed lens model #
tlens = Model(*spt.basics(R), p0=[R])

# Then we compute $\textcolor{Comments}{\theta}$ and $\textcolor{Comments}{\mathsmaller{\tilde{\theta}}}$ #
# We first define a source position ... #
src = (0.01,0.02)

# ... which is modified under the SPT #
hsrc = sm.modified_source_position(src)

# The lensed image positions #
imgs = lens.images(src, omitcore=0.1)

# The SPT-transformed positions #
timgs = tlens.images(hsrc, guess=imgs, profile=True, multi=True)

# We compute the original and #
#  SPT-transformed time delays #
td = lens.time_delays(imgs)
ttd = tlens.time_delays(timgs)
\end{lstlisting}

\section{Impact of the SPT on time delays: empirical estimation}
\label{section:TD}



\begin{figure}
	\centering
	\includegraphics[width=\linewidth]{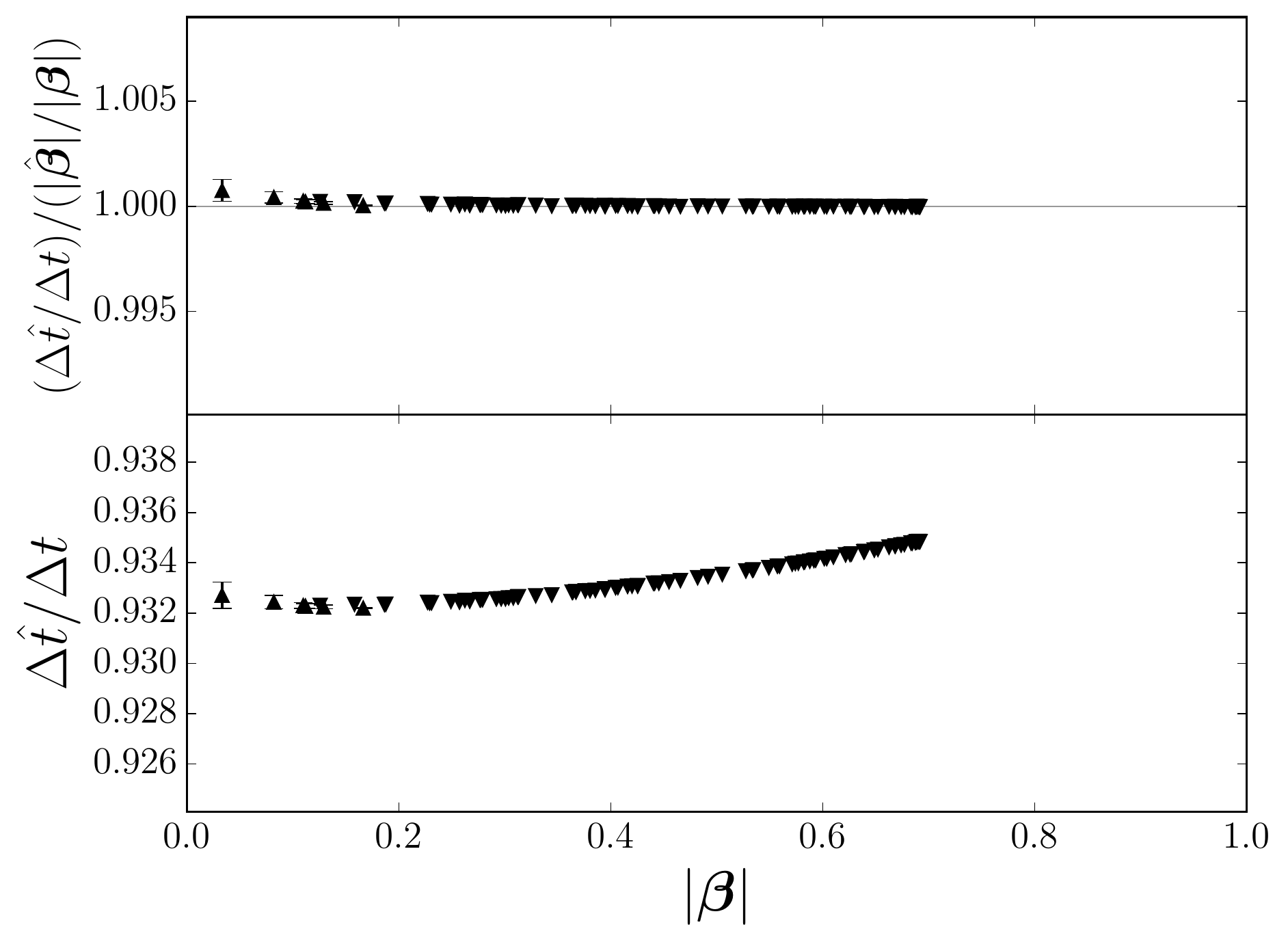}	
     	\caption{	Time delay ratios of image pairs between the composite fiducial model and its SPT-transformed counterpart under 
			a radial stretching with $(f_0,f_2) = (-0.068,0.012)$. 
			\emph{Top:}	$\Delta \hat{t} / \Delta t$ normalized by the ratio $|\bhb|/|\bb|$ are very close to $1$, even in the quadruple image regime, which 
						disagree with the results obtained in \cite{SPT_SS14}.
			\emph{Bottom:}	the impact of the SPT (cleaned from the pure MST with $\lambda = 1 + f_0 = 0.932$) is around only a few tenth of percent.
			The error bars illustrates that the time delay ratios are not conserved in the quadruple image regime, 
			as firstly suggested in \cite{SPT_SS14}.  
			}
       	\label{figure:timedelays_fiducial}       
\end{figure}

\begin{figure}
	\centering
	\includegraphics[width=\linewidth]{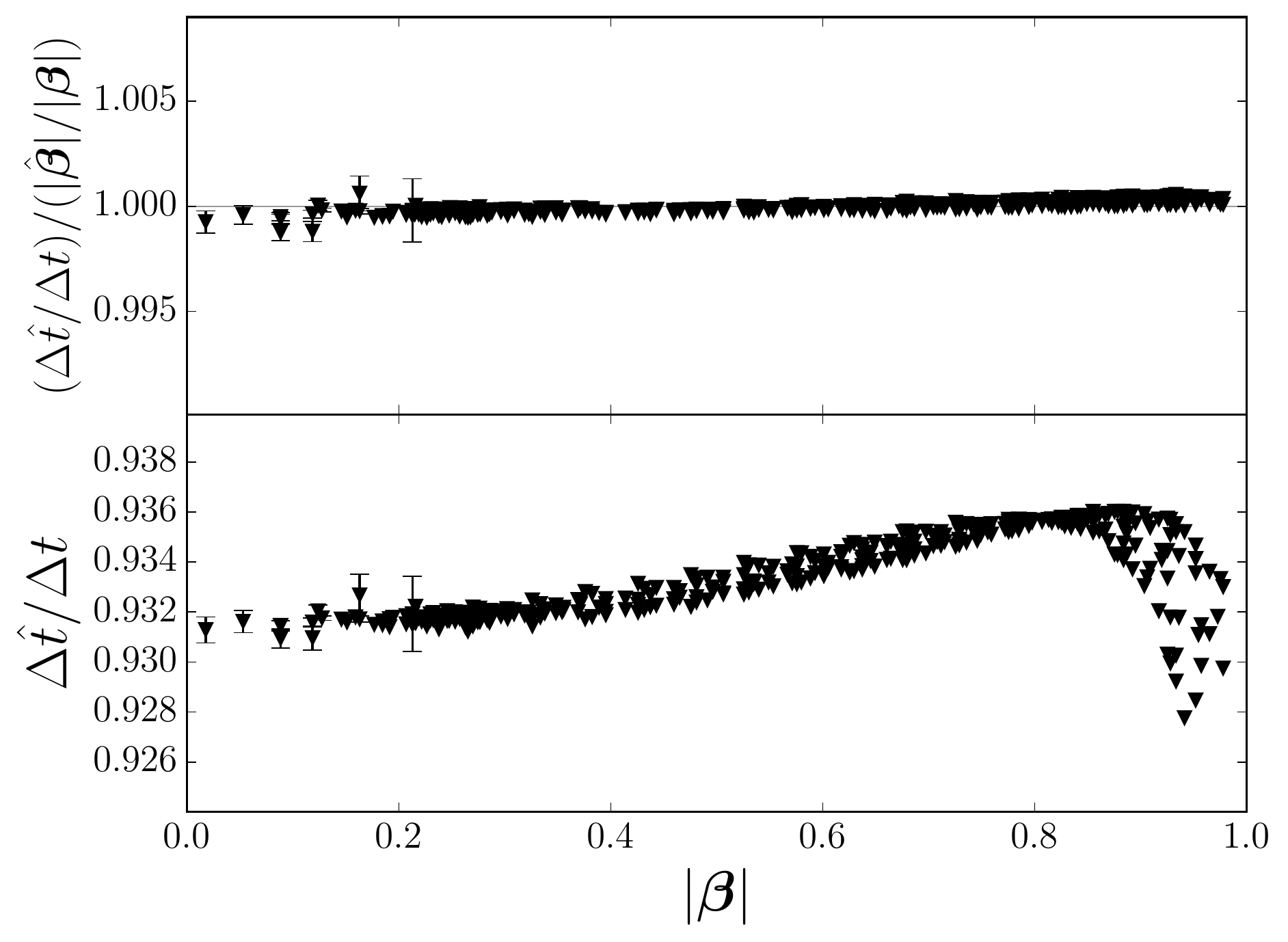}	
     	\caption{	Time delay ratios of image pairs between the composite fiducial model and the quadrupole composed of a SPL.  
			This figure constitutes the corrected version of the figure 4 published in section 4.3 in SS14. The \emph{hat} lensing quantities correspond to the 
			SPL model while the standard notation is used for the composite fiducial model.
			}
       	\label{figure:timedelays_SS14}       
\end{figure}

The first empirical estimation of the impact of the SPT on time delays was presented in \citet{SPT_SS13} and \cite{SPT_SS14}. They showed that a quadrupole model composed of a softened power-law (SPL) 
profile predicts the same lensed image positions (with a $0.004$ arcsec accuracy) as a composite fiducial model (Hernquist profile + generalized NFW + external shear), for a set of sources $\bhb$ and $\bb$, respectively\footnote{
We follow the same notation as in \citet{SPT_SS14}, i.e., we denote the lensing quantities associated with the SPL model with a \emph{hat}, e.g., $\Delta \hat{t}$ for the time delay, whereas no \emph{hat} is used for the composite fiducial model.}. 
We want to stress that the SPL model is not an SPT-generated model from the composite fiducial model but they represent two different models for which the nature of the degeneracy can be approximated by an SPT.
Thus, the set of sources $\bhb$ was obtained independently of $\bb$ by fitting the lensed image positions produced by the fiducial model.
The top panel of the figure 4 in \citet{SPT_SS14} represents $|\bhb|/|\bb|$ as a function of $|\bb|$. For sources located in a disk of radius $0.7\,\text{arcsec}$, the connection between $\bhb$ and $\bb$ is slightly anisotropic and roughly resembles a radial stretching of the form \eqref{deformation_function} with $f_0 = -0.068$ and $f_2 \approx 0.012$.
From the bottom panel of the figure 4 in \citet{SPT_SS14}, they noticed that $(\Delta \hat{t} / \Delta t)/(|\bhb|/|\bb|)$ were almost constant in the double image regime and not conserved in the quadruple image regime. 
They also found that $(\Delta \hat{t} / \Delta t)/(|\bhb|/|\bb|)$ was never smaller than $1.2$, reaching a mean of $1.45$ for the quadruple image configuration of a source located at 
$|\bb| \approx 0.18\,\text{arcsec}$. Thus, this figure shows that the degeneracy between the SPT and fiducial models can affect the inferred value of $H_0$ by an average of $20$\%, up to $45$\% for particular image configurations. 

As a first application of \pythonpackage, we compare these time delay ratios with the ones obtained when we transform the fiducial model under a radial stretching \eqref{deformation_function} with $(f_0,f_2) = (-0.068,0.012)$. 
As shown in Fig.\,\ref{figure:timedelays_fiducial}, we find that the impact of the SPT on time delays is much smaller than predicted in figure 4 in \cite{SPT_SS14}. 
For instance, they found that a source located at $|\bb| = 0.7$ arcsec leads to $\Delta \hat{t}/\Delta t \approx 1.169$ (between the two outer images) whereas we find $\Delta \hat{t}/\Delta t \approx 0.935$, knowing that the major 
contribution comes from an MST with $\lambda = 1 + f_0 = 0.932$. Moreover, the small anisotropic feature of the empirical source mapping $\bhb(\bb)$ alone cannot explain this tension.
%
%
%
%
%

%
The discrepancy between our prediction and the results obtained in \cite{SPT_SS14} are triggered by a minor bug in the public lens modeling code \texttt{lensmodel} \citep[v1.99; ][]{Keeton_MassCatalog_2001} that the authors 
used to compute the time delays. We spotted this bug when we compared the outputs produced with our package \pythonpackage\ and \texttt{lensmodel} for the deflection angle and deflection potential for the SPL model. 
Denoting the logarithm slope of the SPL model as $a$, we found that $2\,\hkp_{\texttt{lensmodel}} = a\,\bn^2 \psi_{\texttt{lensmodel}}$ and $\bha_{\texttt{lensmodel}} = a\,\bn \psi_{\texttt{lensmodel}}$, showing a different 
normalization factor between $\hkp$, $\bha$, and $\hpsi$. This extra normalization factor propagates into the code, leading to a biased value of the time delay $\Delta \hat{t}_{\texttt{lensmodel}}$. It is worth mentioning that, 
for isothermal profiles $(a = 1)$, this extra normalization factor $a$ has no impact on the lensing quantities computed with \texttt{lensmodel}. This might explain why this minor bug has remained unnoticed so far. Nevertheless, the latter has been 
fixed and a corrected version of \texttt{lensmodel} has been immediately released by Chuck Keeton. To obtain the figure 4 in \cite{SPT_SS14}, the SPL model was characterized by the logarithmic three-dimensional slope 
$\gamma' = 2.24$, which is linked to $a$ by the relation $a = 3 - \gamma'$, hence $a = 0.76 \neq 1$, which finally explains the discrepancy mentioned before. 

In Fig.\,\ref{figure:timedelays_SS14}, we show the corrected version of the figure 4 that we produced with our package \pythonpackage. We confirm that the exact same result can now be obtained 
from the corrected version of \texttt{lensmodel}. The normalized time delay ratios $(\Delta \hat{t} / \Delta t)/(|\bhb|/|\bb|)$ plotted against $|\bb|$ (top panel) are very close to $1$. 
Thus, the time delay ratios $(\Delta \hat{t} / \Delta t)$ (bottom panel) closely resembles the source ratios $|\bhb|/|\bb|$ represented in top panel in figure 4. 
This behavior is well understood and is described in detail in the companion paper \citet[][]{SPT_WOS17}.
The impact of the SPT (separated 
from the MST with $1 + f_0 = 0.932$) on time delays now reaches only around $0.32$\% for $|\bb| = 0.7$ arcsec, which is obviously much smaller than the $20\%$ previously found in \cite{SPT_SS14}.
The corrected version now fully agrees with the conclusions drawn from Fig.\,\ref{figure:timedelays_fiducial}. In addition, it confirms that the degeneracy between the SPT and the fiducial models mimics an SPT, 
as firstly established in \citet{SPT_USS17}.

\section{Conclusions}
\label{section:conclusions}

In this paper we have presented the \pythonpackage\ package for analyzing the SPT. \pythonpackage\ relies on several sub-packages, 
the most important of which are \lensing\ to deal with lens model, \sm\ to define an SPT, and \spt\ to provide the SPT-transformed lensing 
quantities defined in previous papers. 
The sub-package \lensing\ is particular in a sense that it can be used independently from any SPT analyze. To some extent, 
it somehwhat offers a \Python-alternative to the public lens modeling code \texttt{lensmodel} with which it shares a lot of functionalities.
\pythonpackage\ implements functionalities for generating lensing quantities produced by SPT-transformed lens models. Thanks to its 
modularity, \pythonpackage\ is also designed to accept both user-defined lens model and SPT. In such a case, \pythonpackage\ constitutes a user friendly 
interface to deal with the SPT. 

As a first application, we have used \pythonpackage\ to explore how a radial stretching may affect the time delay measurements for a fiducial 
model composed of a Hernquist profile + generalized NFW + external shear. 
We found that the impact of the SPT is much smaller than firstly suggested in \citet{SPT_SS14}.  
We have addressed the tension between these results by spotting a minor bug in the public lens modeling 
code \texttt{lensmodel} which was used by
the authors. It resulted in biased values of the deflection potential for the non isothermal SPL model, leading to an overestimated impact of the SPT on the time delays. 
Using the sub-package \lensing, we have produced a corrected version of the figure 4 published in \citet{SPT_SS14}, which now fully agrees with the results presented in this paper. 
As a result, the impact of the SPT on time-delay cosmography might not be as crucial as initially suspected. We address this question in details in the companion paper 
\citet[][]{SPT_WOS17}.

With the next version of \pythonpackage, we plan to include state-of-the-art lens modeling capabilities.
For example, combining stellar dynamics data obtained from spectroscopy of the lens galaxy with lensing measurements has become a standard practice within the strong lensing community.
Furthermore, we still do not have a clear answer to the question: how is the kinematic information of a mass distribution affected under an SPT? 
\citet[][]{SPT_SS13} showed that the fiducial and softened power-law models discussed in Sect. \ref{section:TD} could not be satisfactorily distinguished 
thanks to the measurement of the stellar velocity dispersion $\sigmaP$ with a typical $10\%$ uncertainty. This thus suggests that the use
of $\sigmaP$ may be of limited help for breaking the SPT. 
In a future work, we aim to address this open question with the use of \pythonpackage. In this context, we plan to update the software with a new sub-package fully dedicated to the determination of the 
stellar velocity dispersion associated with an SPT-modified mass profile. 
Finally, \pythonpackage\ is provided not as an ultimate tool for the lens modeling community but primarily as an attractive choice to identify the possible degeneracies that the 
time-delay cosmography may suffer from. Nonetheless we hope that its permanent development will attract more users and will extend its purpose to more than just dealing with the SPT. 





\begin{acknowledgements}
We would like to thank Dominique Sluse and Chuck Keeton for valuable discussions that allowed us to spot and fix the small bug in \texttt{lensmodel}. 
We are very grateful to the anonymous referee for his comments and suggestions that contributed to improving the quality of \pythonpackage\ and this paper.
This work was supported by the Humboldt Research Fellowship for Postdoctoral Researchers. 
\end{acknowledgements}

\bibliographystyle{aa}
\bibliography{bibtex_GL}

\begin{appendix} 


\section{Generalized pseudo-NFW} 
\label{appendix:massmodels}

To our knowledge, no analytical expression of the deflection potential $\psi(\bt)$ for the generalized pseudo-NFW model has ever been published in the literature. For practical purposes, we present
here such an analytical expression, which has been implemented into \pythonpackage. We first recall that the spherical density distribution $\rho(r)$ of the generalized pseudo-NFW model is defined 
as \citep[see the equation 1 in][with $n=3$]{Munoz_CuspModel_2001}
\begin{equation}
	\rho(r) = \frac{\sub{\rho}{s}}{(r/\sub{r}{s})^{\gamma} [1 + (r/\sub{r}{s})^{2}]^{(3-\gamma)/2}} \ , 
	\label{kappa_cuspy}
\end{equation} 
where $\sub{\rho}{s}$ is a characteristic density, $\sub{r}{s}$ the scale radius and $\gamma$ the logarithmic slope of the density profile at small radius.
Up to an additive constant, the axisymmetric deflection potential $\psi(\bt)$ is given by
\begin{equation}
	\label{potential}
	\psi(\bt) = r_{\text{s}}\,\kappa_{\text{s}}\left[\frac{K\left(0,\frac{3-\gamma}{2};\gamma;\frac{|\bt|}{\sub{r}{s}}\right)-K\left(1,0;\gamma;\frac{|\bt|}{\sub{r}{s}}\right)}{\Gamma^{*}(\gamma/2)} - \text{Li}_2\left(-\frac{|\bt|^2}{r_{\text{s}}^2}\right)  \right] ,
\end{equation}
where $\sub{\kappa}{s} = \sub{\rho}{s}\,\sub{r}{s}/\sub{\Sigma}{cr}$, the term $\Gamma^{*}(\upsilon) = \Gamma(\upsilon) / \Gamma(\upsilon-1/2)$ is a particular combination of the gamma function $\Gamma$,  
\begin{equation}
	K(k_0, l ; m; z) = \sum_{k=k_0}^{+\infty} \frac{z^{2(k+l)}}{(k+l)^2} \Gamma^{*}\left[k + l + m/2\right]\, {}_2F_1^{*}\left[k + l, z\right] ,
	\label{Xi}	
\end{equation}
where ${}_2F_1^{*}\left[a, z\right] = {}_2F_1\left[a, a, a + 1, -z^2\right]$ is a particular Gauss hypergeometric function, and the dilogarithm $\text{Li}_2(z)$ can be defined by the series
\begin{equation}
	\label{polylog}
	\text{Li}_2(z) = \sum_{k=1}^{+\infty} \frac{z^k}{k^2}\ .
\end{equation}


%

\section{Proof of the relation \eqref{tilde_psi_simplified}} 
\label{appendix:tildepsi}


As a preamble, the notation adopted here differs from that used in \citet{SPT_USS17}. We use $\bt$ as a position in the lens plane and 
$\bvt$ as the corresponding integration variable for $\bt$. The inverse is partially used in \citet{SPT_USS17}, in particular in the section 3.3 where 
$\tpsi$ and $\bta$ are derived.

Starting with equation 18 in \citet{SPT_USS17}, the deflection potential $\tpsi$ evaluated at the position $\bt$ is given by
\begin{equation}
	\tpsi(\bt) = \langle \tpsi \rangle + 2 \int_{\mathcal{U}} H(\bt;\bvt)\ \hat{\kp}(\bvt)\ \text{d}^2\vartheta - \int_{\partial\mathcal{U}} H(\bt;\bvt)\ \bha \cdot \boldsymbol{n}\ \text{d}s \ ,
	\label{tilde_psi_original}
\end{equation}
where a solution for the Green's function $H$ is analytically known when $\mathcal{U}$ is a disk of radius $R$
\begin{eqnarray}
	H(\bt;\bvt) &=& \frac{1}{4\,\pi} \left[\ln\left(\frac{\left|\bvt-\bt\right|^2}{R^2}\right) + \ln\left(1 - \frac{2\,\bvt \cdot \bt}{R^2} + \frac{|\bvt|^2\,|\bt|^2}{R^4}\right)\right] \nonumber \\
	                 &-& \frac{|\bvt|^2 + |\bt|^2}{4\,\pi\,R^2} \ .
	\label{GreensFunction}
\end{eqnarray}

First, we note that $|\bvt| = R$ for all $\bvt$ on the boundary $\partial \mathcal{U}$, which implies 
\begin{equation}
	1 - \frac{2\,\bvt \cdot \bt}{R^2} + \frac{|\bvt|^2\,|\bt|^2}{R^4} = \frac{|\bt|^2}{R^2} - \frac{2\, \bvt \cdot \bt}{R^2} + \frac{|\bvt|^2}{R^2} = \frac{\left|\bvt-\bt\right|^2}{R^2} \ .
	\label{onboundary}
\end{equation}
Thus, the two logarithm-terms in $H(\bt;\bvt)$ are equal when we consider the line integral.

Secondly, the term $-|\bt|^2/(4 \pi R^2)$ in $H(\bt;\bvt)$ does not depend on $\bvt$, hence contribute neither to the integral over $\mathcal{U}$ nor to the line integral. 
Therefore, Eq.\,\eqref{tilde_psi_original} contains the term 
\begin{equation}
	- \frac{|\bt|^2}{4 \pi R^2} \left(2 \int_{\mathcal{U}} \hat{\kp}(\bvt)\ \text{d}^2\vartheta - \int_{\partial\mathcal{U}} \bha \cdot \boldsymbol{n}\ \text{d}s\right) = 0\ ,
	\label{gaussterm}
\end{equation}
where the equality holds because of $2\, \hkp = \bn \cdot \bha$ and we made use of Gau{\ss} divergence theorem. 
As a result, the term $-|\bt|^2/(4 \pi R^2)$ in $H(\bt;\bvt)$ does not contribute to $\tpsi$.

Finally, combining \eqref{tilde_psi_original}, \eqref{onboundary} and \eqref{gaussterm} leads to the definition of $\tpsi$ given in Eq.\,\eqref{tilde_psi_simplified}.
We note that the same reasoning holds for $\bta$, leading to the Eq.\,\eqref{tilde_alpha_simplified}.



\end{appendix}
\end{document}